\documentclass[12pt,preprint]{adap}
\usepackage{natbib,url,hyperref,enumerate,graphicx}
\usepackage[latin1]{inputenc}   
\usepackage[T1]{fontenc}   
\usepackage{mathptmx} % Times

\usepackage[compact]{titlesec}
\titlespacing{\section}{0pt}{6pt}{4pt}
\titlespacing{\subsection}{0pt}{6pt}{4pt}
\titlespacing{\subsubsection}{0pt}{6pt}{4pt}

\usepackage{enumitem}
\usepackage[usenames,dvipsnames]{color}
\def\bluetext#1{{\color{blue}#1}}

\def\bluetext#1{}
\usepackage[normalem]{ulem}             % allows for strikeout with custom color macros below
\newcommand\redsout{\bgroup\markoverwith{\textcolor{red}{\rule[0.5ex]{2pt}{0.4pt}}}\ULon}
\newcommand\bluesout{\bgroup\markoverwith{\textcolor{blue}{\rule[0.5ex]{2pt}{0.6pt}}}\ULon}
\newcommand\greensout{\bgroup\markoverwith{\textcolor{Green}{\rule[0.5ex]{2pt}{0.6pt}}}\ULon}
\newcommand\magsout{\bgroup\markoverwith{\textcolor{Magenta}{\rule[0.5ex]{2pt}{0.4pt}}}\ULon}
\newcommand\bricksout{\bgroup\markoverwith{\textcolor{Brickred}{\rule[0.5ex]{2pt}{0.4pt}}}\ULon}

\newcommand\reduline{\bgroup\markoverwith{\lower 3.5 pt\hbox{\textcolor{red}{\rule[0.5ex]{2pt}{0.4pt}}}}\ULon}
\newcommand\blueuline{\bgroup\markoverwith{\lower 3.5 pt\textcolor{blue}{\rule[0.5ex]{2pt}{0.4pt}}}\ULon}
\newcommand\greenuline{\bgroup\markoverwith{\lower 3.5 pt\textcolor{Green}{\rule[0.5ex]{2pt}{0.4pt}}}\ULon}
\newcommand\maguline{\bgroup\markoverwith{\lower 3.5 pt\textcolor{Magenta}{\rule[0.5ex]{2pt}{0.4pt}}}\ULon}
\newcommand\brickuline{\bgroup\markoverwith{\lower 3.5 pt\textcolor{Brickred}{\rule[0.5ex]{2pt}{0.4pt}}}\ULon}

\def\indlist#1#2{{\par\noindent \hangindent=4em \hangafter=1 
\noindent\hbox to 3.5em{\hfill {#1}~} {#2}\par}}

\def\onetab{\hbox to 1em{\hss}}
\def\twotab{\hbox to 2em{\hss}}
\def\threetab{\hbox to 3em{\hss}}

\def\lsim{\lower0.3em\hbox{$\,\buildrel <\over\sim\,$}}
\def\gsim{\lower0.3em\hbox{$\,\buildrel >\over\sim\,$}}
\def\rg{\ifmmode{R_{\rm g}}\else{$R_{\rm g}$}\fi}
\def\mbh{\ifmmode{M_{\bullet}}\else{$M_{\bullet}$}\fi}
\def\sol{\ifmmode{_{\odot}}\else{$_{\odot}$}\fi}
\def\msol{\ifmmode{\rm M_{\odot}}\else{${\rm M}_{\odot}$}\fi}
\def\msolyr{\ifmmode{{\rm M_{\odot}\;yr^{-1}}}\else{M$_{\odot}\;$yr$^{-1}$}\fi}
\def\ls{\lower 2pt \hbox{$\;\scriptscriptstyle \buildrel<\over\sim\;$}} 
\def\gs{\lower 2pt \hbox{$\;\scriptscriptstyle \buildrel>\over\sim\;$}} 
\def\kms{\ifmmode{\;{\rm km~s^{-1}}}\else{~km~s$^{-1}$}\fi}
\def\ergs{\ifmmode{\;{\rm erg~s^{-1}}}\else{~erg~s$^{-1}$}\fi}
\def\mic{\ifmmode{\;{\rm \mu m}}\else{$\;\mu$m}\fi}

\def\checkbox{\square\!\!\!\!\!\!\!{\bf X}}

\def \h70{{h_{70}}}
\begin{document}
\thispagestyle{empty}
%%%%%%%%%%%%%%%%%%%%%%%%%%%%%%%%%%%%%%%%%%%%%%%%%%%%%%%%%%%%%%%%%%%%%%%
\begin{center}

{\large\bf Astro2020 Science White Paper}

\bigskip{\Large\bf An Arena for Multi-Messenger Astrophysics:}

\smallskip{\Large\bf Inspiral and Tidal Disruption of White Dwarfs} 

\smallskip{\Large\bf by Massive Black Holes}

\end{center}

\vfill

%\normalsize
%\noindent \textbf{Thematic Areas:} \hspace*{53pt} $\square$ Planetary Systems \hspace*{5pt} $\square$ Star and Planet Formation \hspace*{45pt}\linebreak
%$\checkbox$ Formation and Evolution of Compact Objects \hspace*{31pt} $\square$ Cosmology and Fundamental Physics \linebreak
%$\square$  Stars and Stellar Evolution \hspace*{1pt} $\square$ Resolved Stellar Populations and their Environments \hspace*{40pt} \linebreak
%$\checkbox$    Galaxy Evolution   \hspace*{43pt} $\checkbox$ Multi-Messenger Astronomy and Astrophysics \hspace*{65pt} \linebreak
\noindent{\bf Thematic Areas:}
$$\vbox{\hsize 5truein \halign{# \hfil & #\hfil \cr
$\square$ Planetary Systems & $\square$ Star and Planet Formation \cr
$\checkbox$ Formation and Evolution of Compact Objects & $\square$ Cosmology and Fundamental Physics \cr
$\square$  Stars and Stellar Evolution & $\square$ Resolved Stellar Populations and their Environments \cr
$\checkbox$ Galaxy Evolution & $\checkbox$ Multi-Messenger Astronomy and Astrophysics \cr
}}$$

\vfill
\begin{center}
{\it by}

Michael Eracleous
\\
Department of Astronomy \& Astrophysics and Institute for Gravitation and the Cosmos, The Pennsylvania State University, 525 Davey Lab, University Park, PA 186802, USA
\\
{\tt mxe17@psu.edu}

\bigskip{\it in collaboration with}

Suvi Gezari$^1$, Alberto Sesana$^2$, Tamara Bogdanovi\'c$^3$, Morgan MacLeod$^4$, Nathaniel Roth$^1$, \\ \& Lixin (Jane) Dai$^5$ 

\bigskip{\it and with thanks to}

M. Coleman Miller$^1$ and John Tomsick$^6$ 

for a critical reading of the paper and very helpful comments and suggestions.

\end{center}

\vfill
%\centerline{Submitted on \today}
\centerline{Paper \# 16, submitted on February 18, 2019}
 
\vfill\vbox{\noindent\small
$^1$Department of Astronomy and Joint Space-Science Institute, The University of Maryland, USA \\
$^2$School of Physics and Astronomy, University of Birmingham, UK \\
$^3$Center for Relativistic Astrophysics, School of Physics, Georgia   Institute of Technology, USA \\
$^4$Institute for Theory and Computation, Harvard-Smithsonian Center for Astrophysics, USA \\
$^5$Dark Cosmology Centre, Niels Bohr Institute, University of Copenhagen, Denmark \\
$^6$Space Sciences Laboratory, University of California, Berkeley, USA
}
%
%Lixin (Jane) Dai
%(K{\o}benhavns Universitet Niels Bohr Instituttet, Denmark),

%\medskip\centerline{and the NASA LISA Study Team}

\clearpage
%=============================================
\setcounter{page}{1}

\noindent
{\bf Summary:} The tidal disruption of stars by (super-)massive  black holes in galactic nuclei has been discussed in theoretical terms for about 30 years but only in the past decade have we been able to detect such events in substantial numbers. Thus, we are now starting to carry out observational tests of models for the disruption. We are also formulating expectations for the inspiral and disruption of white dwarfs by ``intermediate-mass'' black holes with masses $\lsim 10^5\;\msol$. Such events are very rich with information and open a new window to intermediate-mass black holes, thought to live in dwarf galaxies and star clusters. They can inform us of the demographics of intermediate-mass black holes, stellar populations and dynamics in their immediate vicinity, and the physics of accretion of hydrogen-deficient material. The {\it combination} of upcoming transient surveys using ground-based, electromagnetic observatories and low-frequency gravitational wave observations is ideal for exploiting tidal disruptions of white dwarfs. The detection rate of gravitational wave signals, optimistically, may reach a few dozen per year in a volume up to $z\approx 0.1$. Gravitational wave observations are particularly useful because they yield the masses of the objects involved and allow determination of the spin of the black hole, affording tests of physical models for black hole formation and growth. They also give us advance warning of the electromagnetic flares by weeks or more. The right computing infrastructure for modern models for the disruption process and event rates will allow us to make the most of the upcoming observing facilities.

\section{Introduction: The Tidal Disruption of a Star by a Massive Black Hole} \label{sec:intro}

When a star gets close enough to a (super)massive black hole so that the tidal acceleration from the black hole exceeds its surface gravity it is stretched along its direction of motion and compressed perpendicular to it, it is heated, and eventually gets disrupted \citep[e.g.][]{rees88,Ulmer99,Kobayashi04,Guillochon09}.
The critical distance from the black hole at which disruption becomes possible is known as the tidal disruption radius and can be cast in terms of the basic parameters of the system: $R_t \approx (\mbh/m_*)^{1/3}\, r_*$, where $m_*$ and $r_*$ are the mass and radius of the star, and $\mbh$ is the mass of the black hole.  The post-disruption debris returns to the black hole a dynamical time later (about a month later, for a $10^6\;\msol$ black hole and a $1\;\msol$ star) and is accreted by the black hole leading to a bright flare. Thus, the disruption of a solar-type star by a $10^7\;\msol$ black hole is easily detectable to a redshift of 0.1 (just under 0.5 Gpc) because of the initially high accretion rate, which exceeds the Eddington limit of black holes of this mass. 

%A brief flash due to tidal heating of the star during the first pericenter passage and a (relatively weak) burst of gravitational waves precede. 

Tidal disruption flares were predicted several decades ago \citep[see][]{rees88} but since their rates are of the order of $10^{-5}$--$10^{-4}$ per galaxy per year, they were only detected in substantial numbers fairly recently, with the advent of systematic surveys for transients, such as Pan-STARRS1, the Palomar Transient Factory, and ASAS-SN. The classic observational signature of tidal disruption events is an accretion-powered flare in the soft X-ray/UV/optical bands with a very sharp rise over a few days and a gradual decline over a few weeks that typically follows a power law, dictated by the evolution of the accretion rate of the post-disruption debris.
The couple of dozen events discovered so far have allowed estimates of their rate that are in general agreement with theoretical predictions \citep{vanVelzen18}. We have also been able to study their phenomenology by characterizing the shapes of their light curves and the properties of their spectra. The Large Synoptic Survey Telescope (LSST) will find thousands of tidal disruption events per year once it becomes operational in the next decade. With many events we will be able to probe the demographics of supermassive black holes in the centers of galaxies, constrain dynamical models for stars in their vicinity, and test models for the accretion flows that give rise to the flares. 

%---------------------------------------

\section{Getting to the Point: Tidal Disruption of White Dwarfs}
\label{sec:compact}

The disruption of white dwarfs proceeds in a similar fashion as the disruption of solar-type stars, but it requires less massive black holes in order to be observable. Since the tidal disruption radius scales as $R_t\propto \mbh^{1/3}$ while the radius of the event horizon scales as $R_S\propto\mbh$, for a given combination of a star's mass and radius there is a maximum black hole mass above which the tidal disruption
is unobservable because it happens within the black hole's event horizon. This limit is $\sim$$10^8\;\msol$ for a solar-type star and a non-rotating black hole, and it decreases for denser stars. A 0.5~\msol\ white dwarf is disrupted outside of the event horizon of a non-rotating black hole with $\mbh \lsim 2\times 10^5 \;\msol$ ($\lsim 10^6\;\msol$, for a rapidly-spinning black hole and/or a low-mass white dwarf). If $\mbh \lsim 4\times 10^4 \;\msol$, the white dwarf is disrupted outside the innermost stable circular orbit, and the debris may form an accretion disk around the black hole.

Depending on the orbit and mass of the white dwarf, various outcomes are possible, each interesting in its own right. For example, the close flyby of the white dwarf can generate a burst of gravitational waves detectable by a space-based observatory (e.g., LISA, the {\it Light Interferometer Space Antenna}), if it occurs within a Milky Way satellite \citep{rosswog09,anninos18}. For very close encounters, the tidal compression of the white dwarf may lead to detonation and a flare that resembles a Type~Ia supernova \citep{rosswog08,MacLeod16}. If such an event occurred in a Milky Way globular cluster, it might produce a detectable neutrino flux \citep[ala type Ia supernovae; see][]{kunugise07,wright16,seitenzahl15}. So, the disruption of a white dwarf is a rich source of information on the physical processes involved.

%---------------------------------------

\section{The Main Point: White Dwarfs Spiraling into  Massive Black Holes}
\label{sec:special}

A particularly interesting type of tidal disruption event is the disruption of a white dwarf in a {\it bound} orbit around a moderately massive black hole, discussed in some detail by \citet{sesana08} and others. Such events are likely the most common type of ``extreme mass-ratio inspirals'' (EMRIs), in which the white dwarf is effectively a test particle that probes the spacetime around the black hole as it spirals in.
The evolution of a fiducial system comprising a $0.5\;\msol$ white dwarf in a tight, bound orbit around a $10^5\;\msol$ black hole proceeds as follows. The white dwarf is captured in a bound orbit around the black hole either directly or after the breakup of a bound stellar binary of which it was a member \citep[e.g.][and references therein]{MacLeod14,MacLeod16}. The initial bound orbit is likely to be very eccentric with a pericenter distance $\sim 10$--$15\;\rg$, where $\rg \equiv G\mbh/c^2$ is the gravitational radius of the black hole ($\rg\approx 1.5\times 10^6\;{\rm km}\approx 2\;R_\odot$ in this case), and an orbital period of order a few minutes. The orbit decays gradually by the emission of gravitational waves (with initial frequency $f_{\rm gw}\sim 10\;$mHz) and/or by the tidal excitation of oscillations in the white dwarf \citep[e.g.][]{Bogdanovic14,MacLeod14}.  A gravitational-wave observatory can detect this phase of the inspiral out to distances of 250--450~Mpc \citep[depending on the masses and orbital eccentricity;][the estimate holds for the currently advertised LISA too]{sesana08}. After about 4 years, the white dwarf approaches its tidal disruption radius ($R_t \approx 5 \rg$ and $f_{\rm gw}\sim 80\;$mHz, in this particular case) and begins to lose mass because of its tidal deformation. The accretion of this material by the black hole now makes the system visible via electromagnetic waves \citep[with a luminosity of $10^{43}\;{\rm erg\;s}^{-1}$;][]{zalamea10}. The orbital evolution and attendant gravitational wave signal are influenced by the mass transfer, offering additional indicators of the properties of the system, most notably the mass \citep{dai13}. If the white dwarf is in a very eccentric orbit, mass loss and accretion occur during pericenter passage, making the electromagnetic signal periodic. This phase lasts for days to weeks before the white dwarf is completely disrupted, as depicted in Figure~1, and accretion of the debris leads to a bright electromagnetic flare. 

\bigskip
\centerline{
\includegraphics[scale=0.293]{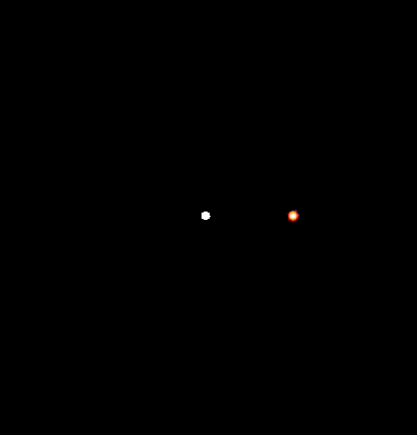}
\includegraphics[scale=0.3]{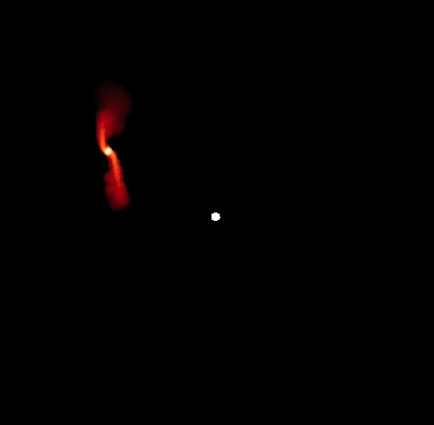}
\includegraphics[scale=0.3]{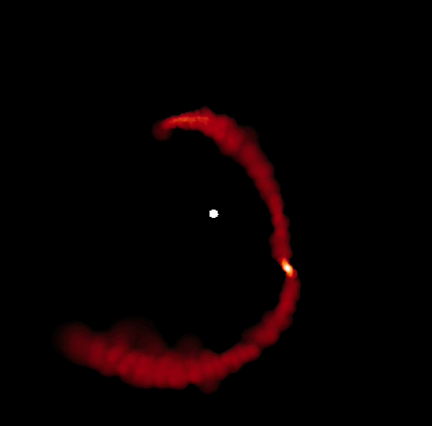}
}
\medskip
\centerline{\vbox{\hsize=5.5truein\noindent
\textbf{\textsc{ Figure 1.--}} {\small Three frames from a simulation of the disruption of a $0.6\;\msol$ white dwarf in a nearly circular orbit around a $10^4\;\msol$ black hole (by D. Clausen, unpublished, used with permission). The color encodes column density. The orbital period is 16~s and the three frames span less than one orbital cycle (the white dwarf moves counter-clockwise). Each frame is $5\times 10^{10}\;$cm (or $34\;R_{\rm g}$) on the side.}}}
\medskip

The rate of white dwarf inspirals is $\sim 100~{\rm yr^{-1}~Gpc^{-3}}$ in dwarf galaxies and $\sim 1~{\rm yr^{-1}~Gpc^{-3}}$ in globular clusters \citep[assuming, optimistically, that {\it all} dwarf galaxies and globular clusters host massive black holes and updating estimates by][]{MacLeod14,MacLeod16}, but they are uncertain by up to two orders of magnitude. In comparison, the rate of disruption of white dwarfs in unbound orbits is 10 times higher and the corresponding rate for main sequence stars is  1,000 times higher. 

The combined gravitational wave and electromgannetic signals from many inspirals and the ensuing disruptions can teach us a great deal \citep[see][]{amaro07,sesana08}.

%\begin{enumerate}[noitemsep,topsep=0in,label=({\alph*})]

%\item 
\noindent{\bf (a)}
\textbf{\textsc{Black hole demographics:}} The events are signposts of  moderately massive black holes (i.e., $\mbh \lsim 10^5 \;\msol$) and allow us to explore their demographics and the properties of their host galaxies or star clusters. Black holes in this mass range are the sought after but elusive ``seeds'' invoked in models of the co-evolution of galaxies and their central black holes \citep[e.g.,][]{volonteri12}. A few dozen candidates have been found in dwarf galaxies \citep[e.g.,][]{Reines13,Moran14,Baldassare16}, and some in globular clusters, but determining their masses has been a challenge. 
%Similarly, black holes with masses $\mbh \lsim 10^3 \;\msol$ may be present in the centers of globular clusters, but the candidates proposed so far \citep[e.g.,][]{Gebhardt05,maccarone07,Kiziltan17,Perera17,Lin18} still await confirmation. 

%\item 
\noindent{\bf (b)}
\textbf{\textsc{Stellar populations and dynamics:}} The event rate provides valuable constraints on the stellar populations around the black holes
%and globular clusters
and on the dynamical processes that lead to stellar captures. The rate of white dwarf captures depends on various assumptions, including the mass spectrum of black holes, the probability that a dwarf galaxy has a central massive black hole at all, the stellar orbits near the center of the galaxy, 
%mass segregation, 
and many others. The observed tidal disruption rates and gravitational wave signals will, then, eliminate a substantial subset of the available models and narrow the range of reasonable assumptions. Moreover, the optical/ultraviolet spectrum of the electromagnetic flare tells us about the composition of the white dwarf \citep[e.g.][]{Clausen11,Clausen12}, which can be connected to its mass.

%\item 
\noindent{\bf (c)}
\textbf{\textsc{Fundamental system parameters:}} The gravitational wave signal gives the masses of the two objects. Since a very large number of orbital cycles can be observed during the inspiral, it is possible to determine the white dwarf and black hole masses to 1 part in $10^3$ or better. Since the white dwarf spends a lot of its time in close proximity to the black hole (from 10 to 5$\;\rg$, or $< 5\;\rg$ for more massive white dwarfs), the gravitational wave signal allows us to map the spacetime around the black hole and determine the magnitude and direction of the black hole spin. Hence we can test scenarios for the formation and growth of black holes in this mass range via their predictions for the distribution of spins \citep[see][]{King08,Stone17}

%\item 
\noindent{\bf (d)}
\textbf{\textsc{Accretion physics:}} The time evolution of the electromagnetic flare conveys information about the physics of the accretion of the hydrogen-defficient debris (including the formation of jets and other outflows). The gravitational wave signal will have already given us the masses of the two objects involved {\it and} the black hole spin, which are invaluable but could not be obtained otherwise. Thus, we will be able to test models for super-Eddington accretion flows \citep[e.g.,][]{Jiang14,Jiang17,dai18} over a very wide range of Eddington ratios and the transition from super- to sub-Eddington flows (because of the lower black hole masses and shorter time scales).
  
%\item 
\noindent{\bf (e)}
\textbf{\textsc{Cosmology:}} Electromagnetic plus gravitational wave observations of many events yield a distance-redshift relation that is independent of the well-known uncertainties of the cosmological distance ladder and a value of the Hubble constant. White dwarf inspirals fill the gap in redshift between stellar mass and supermassive binary black hole mergers \citep[see][]{tamanini16}. 
%The implications of this measurement for cosmology are manifest.
  
%\end{enumerate}

%---------------------------------------

\section{Future/Upcoming Facilities and Capabilities That Will Make a Difference}
\label{sec:facilities}

Current wide-field optical and X-ray surveys are detecting tidal disruption events at a rate of $\sim 2$ per year. The distribution of black hole masses for currently known tidal disruption events peaks at $\sim 10^6$--$10^7\;\msol$ \citep{Wevers17,vanVelzen18}. The latter result is not surprising since black holes in this mass range are the most common of those accessible to current surveys. Moreover, in view of the expected relative disruption rates noted in \S\ref{sec:special}, it is rather unlikely that any of the disruptions observed so far involve white dwarfs. But, in the next decade, the survey capabilities in the optical time domain will dramatically improve with the recent commissioning of the Zwicky Transient Facility (ZTF) and the advent LSST in 2022. The large samples expected from ZTF of tens per year \citep{Hung17} and LSST of several thousands per year \citep{vanVelzen11} will have the depth to probe much further down the black hole mass function, reaching masses below $10^5\;\msol$. Therefore, one can reasonably expect that the LSST will discover at least a dozen white dwarf disruptions per year. The survey capabilities of ZTF and LSST will also allow detections of tidal disruption events in the very local universe. For example, in a dedicated search for tidal disruption events with the intermediate Palomar Transient Factory (iPTF), a fainter event that evolved very fast was discovered at a distance of only 67~Mpc \citep[iPTF 16fnl:][]{Blagorodnova17}. This discovery highlights the prospect of discovering white dwarf inspirals very nearby.  

A low-frequency gravitational wave observatory like LISA (due to launch in $\sim 15$ years) has the ideal capabilities to observe white dwarf inspirals. It is most sensitive at $f_{\rm gw}\sim10\;$mHz, which corresponds to the longest phase of the inspiral. Moreover, LISA would be able to follow the inspiral to almost $f_{\rm gw}\sim100\;$mHz, i.e., no more than a year before the disruption (depending on the white dwarf and black hole masses). The volumetric rate of white dwarf inspirals and the volume accessible by LISA imply, optimistically, that the dozen electromagnetic flares from white dwarf inspirals and ensuing disruptions detected per year by LSST will have their gravitational wave signals detected a few years earlier. Moreover, since the inspiral time of a white dwarf into a $10^5\;\msol$ black hole is of order years, an even larger number of such systems will be detectable by LISA at different stages of the inspiral. Therefore, models for event rates can be readily tested and there will be many opportunities for detailed case studies.

A decade ago models of the disruption of a solar-type star by a supermassive black hole that neglected relativity and a variety of other effects were sufficient for painting a basic picture of the process and for interpreting the early observations. But the discoveries by the PanSTARRS and ASAS-SN surveys, among others, underscore the need for modern, relativistic, magnetohydrodynamic models that include the interaction of radiation and matter in the post-disruption flare and can describe super-Eddington accretion flows. Such models should follow the evolution of the debris over long time scales in order to capture the formation of an accretion disk \citep[e.g.,][are steps in the right direction]{shiokawa15,anninos18,dai18}. Predictions of event rates from detailed stellar dynamics calculations must also be updated to let us take full advantage of the {\it observed} event rates. Each of these computational problems is challenging in its own right but they are all pertinent to tidal disruptions. It is therefore necessary to invest in the development of computational infrastructure that will be readily available to the theoretical astrophysics community. Answering the questions laid out above requires not only a synergy between observing facilities but also a synergy between theory and observations. 

%\end{document}

%%%%%%%%%%%%%%%%%%%%%%%%%%%%%%%%%%%%%%%%%%%%%%%%%%%%%%%%%%%%%%%%%%%%%%%
%%%  BIBLIOGRAPHY
%%%%%%%%%%%%%%%%%%%%%%%%%%%%%%%%%%%%%%%%%%%%%%%%%%%%%%%%%%%%%%%%%%%%%%%

\clearpage
\label{sec:references}
\bibliographystyle{jponew_3auth_etal}
\bibliography{references.bib}

\end{document}